\documentclass {article}

\title{Boltzmann entropy and the microcanonical ensemble}
\author{J. Naudts\thanks {Email: jan.naudts@ua.ac.be}\\
           \small Universiteit Antwerpen -
           Universiteitsplein 1, 2610 Antwerpen, Belgium
}
\date {}

\input psfig.sty

\def\upd{{\rm d}}
\newcommand\Name[1]{#1,}
\newcommand\REVIEW [4]{#1{\bf #2}, #4 (#3).}

\begin{document}

\maketitle

\begin{abstract}
Boltzmann's entropy is slightly modified to make it suitable for discussing phase transitions
in finite systems. As an example it is shown that the pendulum undergoes a second
order phase transition when passing from a vibrational to a rotating state.
\end{abstract}

\section{Introduction}

The Boltzmann-Gibbs formalism of statistical mechanics is heavily based
on the equivalence of ensembles, which for many systems holds in the thermodynamic
limit. It is not adequate to describe the thermodynamic behaviour of small systems.
In particular, phase transitions can only occur in the thermodynamic limit of the
canonical ensemble. Recently \cite {ZGXZ87,GD90,HA94,PH95,GV00,GDH01,BH04},
the use of the microcanonical ensemble has been
advocated to describe phase transitions in small systems. The present Letter supports
this approach, but argues that one of its premises should be slightly modified.
Indeed, it is tradition to use Boltzmann's entropy
\begin {eqnarray}
S_{\mathrm{B}}(E)=k_{\mathrm{B}}\ln(\epsilon \omega(E)),
\label {Bent}
\end {eqnarray}
with $\epsilon$ a small but fixed energy and with $\omega(E)$ the density of states,
as the obvious definition of thermodynamic entropy of isolated systems.
Alternatively (see \cite {PHT85} and the references cited there),
the density of states $\epsilon \omega(E)$ in (\ref {Bent})
can be replaced by the integrated density of states $\Omega(E)$.
The proposal of the present Letter is to replace it by
a related quantity $\Theta(E)$, solution of the non-linear differential equation
\begin {eqnarray}
\frac {\Theta'(E)^2}{\Theta(E)\Theta''(E)}-1=1-\frac {\omega(E)\omega''(E)}{\omega'(E)^2}.
\label {difeq}
\end {eqnarray}
The resulting entropy is denoted $S_\Theta(E)$
\begin {eqnarray}
S_\Theta(E)=k_{\mathrm{B}}\ln(\Theta(E)).
\label {Thent}
\end {eqnarray}
A conceptual argument supporting this modification is given below.
Practical consequences for macroscopic systems are neglegible.
But for small systems the consequences are important.

\section{Equipartition}

According to Boltzmann the equilibrium state of an isolated system
is the macrostate that can be realized in the largest number of ways
\cite {UJ04}. All accessible states have the same probability. This can be taken
as the definition of the microcanonical ensemble.
The notion of entropy enters when two or more isolated systems are compared.
The density of states in the combined system is the product of density
of states of each of the composing systems. On the other hand,
thermodynamic entropy must be the sum of the entropies of the subsystems.
The standard argument is then that entropy necessarily is proportional
to the logarithm of the density of states. This results in Boltzmann's entropy.
Problem with this reasoning is that
the compound system still consists of two unrelated systems, e.g.~physically
separated by a wall. Removal of this wall would affect entropy.
If the wall is not removed then the compound system is in equilibrium,
but this is the equilibrium of a microcanonical ensemble constraint to the
requirement that the different subsystems, labeled with index $j$, have fixed energies $E_j$.
Not all possible ways of distributing the total energy $E=\sum_jE_j$ over
the subsystems are realized.
Hence it is not in equilibrium in the unconstraint ensemble.
The usual way to overcome this problem is by means of equivalence of ensembles.
However, this works only for macroscopic systems,
if it works at all. Indeed, the essence of the microcanonical approach
to phase transitions in finite systems is precisely that the equivalence
of ensembles can break down in presence of phase transitions. When this happens
then the microcanonical ensemble is the one that decribes nature, not the
canonical ensemble.
Several conclusions can be drawn from this discussion.
In the first place it shows that the standard argument for associating
Boltzmann's entropy with the microcanonical ensemble is not justified
in case of small systems. It also shows that the possibility of
phase separation should be taken into account because it modifies the
accessible part of phase space.

A different type of reasoning is now followed. It starts from the assumption that
it is possible to compare isolated systems and that it is meaningful to state that
they all have the same microcanonical temperature.
Note that energy is the control parameter of the microcanonical ensemble, not
temperature. Therefore the argument is restricted to simple systems
for which energy and temperature are uniquely related.
Initially, put all the isolated
subsystems in their ground state. Next distribute small amounts of energy
d$E$ to the system in such a way that all degrees of freedom of the system are treated
in the same manner. To this purpose, introduce an effective number of
degrees of freedom $m_j(E_j)$ which may depend on the energy $E_j$ of subsystem $j$.
Then the amount of energy received by subsystem $j$
is proportional to $m_j(E_j)$.
Let us postulate that this procedure of distributing energy over the
subsystems keeps them all at the same temperature $T=T(E)$.
For each of them is then
\begin {eqnarray}
C_j(E_j)\equiv
\frac {\upd E_j}{\upd T}=C(E)\frac {m_j(E_j)}{\sum_k m_k(E_k)}
\end {eqnarray}
with $C(E)$ the heat capacity of the total system.
The obvious solution of this equation is
\begin {eqnarray}
C_j(E_j)=k_{\mathrm{B}} m(E_j).
\end {eqnarray}
Integration of $1/C_j(E_j)$ then gives temperature $T$ as a function of energy $E_j$.

\section{Modelling heat capacity}

Let us next try to relate the effective number of
degrees of freedom $m(E)$ to the density of states $\omega(E)$. A suitable definition is
\begin {eqnarray}
m(E)=1+\frac {\omega'(E)^2}{\omega'(E)^2-\omega(E)\omega''(E)}.
\label {degfree}
\end {eqnarray}
For a system with Hamiltonian $H$ the density of states equals
\begin {eqnarray}
\omega(E)=\int_\Gamma\upd \gamma\,\delta(E-H(\gamma)).
\end {eqnarray}
In case of the harmonic oscillator is $\omega'(E)=0$, which makes (\ref {degfree}) badly defined.
However, if $\omega(E)=AE^\nu$, with $A$ constant, then $m(E)=1+\nu$ follows.
Hence, taking the limit $\nu=0$ one concludes that for a constant density of states
one has $m(E)=1$, as expected. This implies
$E=k_{\mathrm{B}}T$, which is the canonical result.
In case of the ideal gas is $\omega(E)\sim E^{3n/2-1}$.
This implies $m(E)=3n/2$ and hence $E=(3n/2)k_{\mathrm{B}}T$ which is again the canonical result.
One concludes that (\ref {degfree}) gives physical results in both cases.

Note that one has
\begin {eqnarray}
m(E)=1-\frac {(S'_{\mathrm{B}}(E))^2}{k_{\mathrm{B}}S''_{\mathrm{B}}(E)}.
\label {ms}
\end {eqnarray}
If the system is macroscopic then the constant 1 in this expression may be omitted,
in which case integration can be done easily. The result is that in that case
Boltzmann's entropy is the thermodynamic entropy, satisfying the thermodynamic relation
\begin {eqnarray}
\frac 1T=\frac {\upd S}{\upd E}.
\label {thermtemp}
\end {eqnarray}
In the general case Boltzmann's entropy is not the thermodynamic entropy.
The proposal is to replace it by $S_\Theta(E)$ with $\Theta(E)$ chosen in such a way
that (\ref {thermtemp}) is satisfied
and that the heat capacity $C(E)$ equals $k_{\mathrm{B}}m(E)$, with $m(E)$ given by (\ref {ms}).
This implies that $\Theta(E)$ must satisfy
\begin {eqnarray}
\frac 1{m(E)}=1-\frac {\Theta(E)\Theta''(E)}{(\Theta'(E))^2}.
\end {eqnarray}
In combination with (\ref {degfree}) this gives (\ref {difeq}).

\section{Discussion}

Let us investigate some properties of $S_\Theta(E)$. A short calculation yields
\begin {eqnarray}
\frac {\upd^2\,}{\upd E^2}S_\Theta(E)
&=&-k_{\mathrm{B}}\left(\frac {\Theta'(E)}{\Theta(E)}\right)^2\frac 1{m(E)}.
\end {eqnarray}
Hence the condition for $S_\Theta(E)$ to be concave is that $m(E)>0$.
On the other hand, it is clear from (\ref {ms}) that the condition for
Boltzmann's entropy to be concave is $m(E)\ge 1$. One concludes that
some of the systems that are unstable w.r.t.~Boltzmann's entropy
are still stable when analysed with $S_\Theta(E)$. In particular, negative
heat capacities, often discussed in the context of finite systems
\cite {GFFGBCM01,GFGBCSM02,GC02,RGM03,GCRR03,GDH04},
are less likely to be negative when using $S_\Theta(E)$.

Of practical importance is that (\ref {difeq}) can be solved in a systematic manner.
Introduce a function
\begin {eqnarray}
f_{\mathrm{B}}(E)=k_{\mathrm{B}}\frac {\upd\,}{\upd E}\frac 1{S'_{\mathrm{B}}(E)}
=1-\frac {\omega(E)\omega''(E)}{\omega'(E)^2}.
\end {eqnarray}
The function $f_\Theta(E)$ is defined in a similar way. Then (\ref {difeq}) can be written as
\begin {eqnarray}
k_{\mathrm{B}}\frac {\upd\,}{\upd E}\frac 1{S'_\Theta(E)}
=f_\Theta(E)=\frac {f_{\mathrm{B}}(E)}{f_{\mathrm{B}}(E)+1}
=\frac {\omega(E)\omega''(E)-\omega'(E)^2}{\omega(E)\omega''(E)-2\omega'(E)^2}.
\end {eqnarray}
By integration of $f_\Theta(E)$ one obtains the inverse of $S'_\Theta(E)$.
Inverting and integrating again one obtains $S_\Theta(E)$.

Finally, note that the two entropies $S_{\mathrm{B}}(E)$ and $S_\Theta(E)$ coincide only when the
density of states is an exponential function $\omega(E)=A\exp(bE)$.
In that case the effective number of degrees of freedom is infinite.

\section{Constant heat capacity}

Of interest is the situation where $m(E)$ is
constant. E.g, if $\omega(E)=AE^\nu$, then
one has $m(E)=1/(1-\nu)$. This case includes the harmonic
oscillator as well as the ideal gas. However, there is a larger class
of systems with constant $m(E)$.
Note that from (\ref  {degfree}) follows
\begin {eqnarray}
\frac {\omega(E)\omega''(E)}{\omega'(E)^2}=1-\frac {1}{m-1}\equiv q.
\end {eqnarray}
Solutions of this equation are of the form $\omega(E)=A\exp_q(bE)$,
involving the deformed exponential functions \cite {TC94,NJ02}
\begin {eqnarray}
\exp_q(x)=\big[1+(1-q)x]_+^{1/(1-q)}
\end {eqnarray}
(the notation $[x]_+=\max\{0,x\}$ is used). The corresponding
$\Theta$-entropy (determined up to a constant term which is fixed by choosing an arbitrary
positive energy $\epsilon$) is
\begin {eqnarray}
S_\Theta(E)=k_{\mathrm{B}} m\ln\frac {E-E_g}{m\epsilon}.
\end {eqnarray}
Here, $E_g$ is the ground state energy of the system.
Temperature $T$, calculated using (\ref {thermtemp}), satisfies
\begin {eqnarray}
k_{\mathrm{B}}T=\frac {E-E_g}m.
\end {eqnarray}
One concludes that any microcanonical system described by a density of states of the form $\omega(E)=A\exp_q(bE)$
satisfies the equipartition law.

\section{The pendulum}

An interesting example of a finite system with phase transition is the pendulum with Hamiltonian
\begin {eqnarray}
H=\frac 1{2I}p^2-\alpha^2I\cos(\phi).
\end {eqnarray}
For low energy $-\alpha^2I<E<\alpha^2I$ the motion is oscillatory.
At large energy $E>\alpha^2I$ it rotates
in one of the two possible directions. A correct understanding of the thermostatistics of
this model is highly relevant because rotational motion of ions and molecules, or part of
molecules, occurs frequently in all states of matter.
See e.g.~\cite {SJN79}.

\begin{figure}
\centerline{\psfig{figure=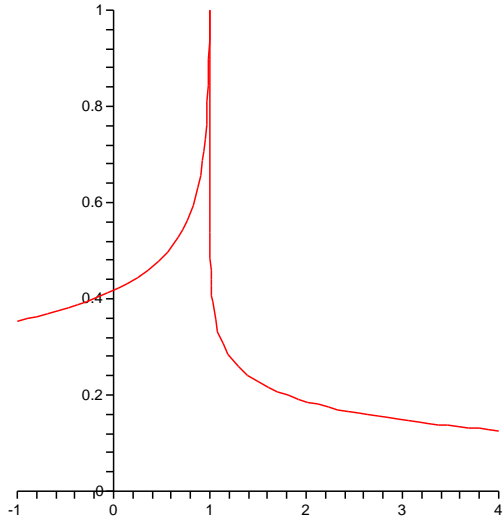,width=8cm}}
\caption{Density of states $\omega_0(u)$ of the pendulum
as a function of energy in reduced units.}
\label{f.1}
\end{figure}

The density of states $\omega(E)$ can be written as
\begin {eqnarray}
\omega(E)=\frac {2\sqrt 2}{\epsilon\alpha}\,\omega_0(E/\alpha^2I).
\end {eqnarray}
with $\epsilon$ a constant, inserted for dimensional reasons, and
with $\omega_0(u)$ given by
\begin {eqnarray}
\omega_0(u)
&=&\frac 1{2\pi}\int_0^1\upd x\,\frac 1{\sqrt x}\frac 1{\sqrt {1-x}}\frac 1{\sqrt{1-u+(1+u)x}}
\qquad \mbox{ if }\quad -1<u<1,\cr
&=&\frac 1{4\pi}\int_{-1}^1\upd x\,\frac 1{\sqrt{1-x^2}}\frac 1{\sqrt{x+u}}
\qquad \mbox{ if }\quad 1<u.
\end {eqnarray}
See fig.~\ref{f.1}. Note that a factor $1/2$ has been inserted in case $u>1$
to take into account that the pendulum is rotating either clockwise or anti-clockwise.

\begin{figure}
\centerline{\psfig{figure=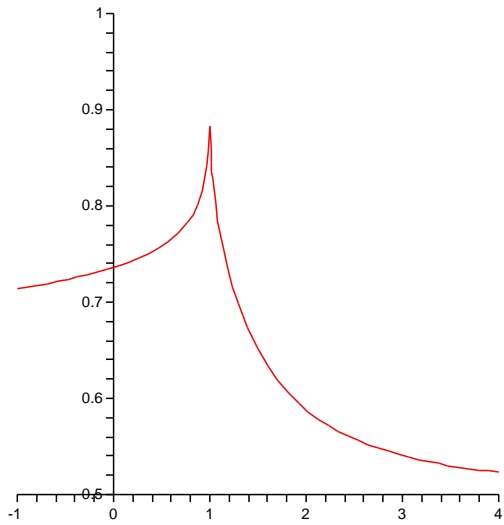,width=8cm}}
\caption{Number of degrees of freedom $m(u)$ of the pendulum
as a function of energy in reduced units.}
\label{f.2}
\end{figure}

The density of states $\omega(E)$ diverges at $E=\alpha^2I$. It is log-convex on both intervals
$-\alpha^2I<E<\alpha^2I$ and $\alpha^2I<E$, rather than log-concave. Hence, Boltzmann's entropy cannot be
used to analyse this model. The effective number of degrees of freedom $m(E)$
has the value 1/2 at large values of $E$, as expected. However, it does not start
with the value 1 at $E=-\alpha^2I$, but with $m(E)\simeq 0.72$. This is due to
anharmonicity of the oscillatory motion.
The heat capacity $C(E)=k_{\mathrm{B}}m(E)$ has a shape typical for a second order
phase transition at $E=\alpha^2 I$.
See fig.~\ref{f.2}.

\section{Conclusions}

Arguments have been given why Boltzmann's entropy is not suited to discuss
phase transitions in small systems and should be slightly modified.
The proposed replacement is given by (\ref {Thent}). It involves
a function $\Theta(E)$, related to the density of states $\omega(E)$
and obtained by solving (\ref {difeq}). It is based on the notion
of effective number of degrees of freedom $m(E)$ and the principle that,
if two systems with equal $m(E)$ are heated up in the same manner, then
they have at all times the same temperature. Crucial point is relation
(\ref {degfree}) between $m(E)$ and the density of states $\omega(E)$.
Omitting the constant term '$+1$' in this expression leads to
Boltzmann's entropy. Hence, if the number of degrees of freedom
is macroscopic then the proposed entropy (\ref {Thent}) coincides
with Boltzmann's entropy. But for small systems one can expect important
differences.

To illustrate this point the classical pendulum has been discussed.
In the present approach it exhibits a second order phase transition between
vibrations and rotations. It cannot be treated using Boltzmann's entropy
because the density of states is piecewise convex instead of concave.
In the approach of \cite {PHT85} it undergoes a first order transition
because heat capacity is negative in a range of energies (this calculation
has not been included but can be done straightforwardly).
Negative heat capacities have been discussed at many places in the literature,
in both theoretical and experimental papers. It is out of scope of the
present Letter to reanalyse all those applications. But it is clear that
this has to be done.

The main omission of the present Letter is a discussion of the quantum mechanical
microcanonical ensemble. In the quantum case the density of states $g(E)$ counts
the number of eigenstates with energy in the range
from $E$ to $ E+\epsilon$, for some fixed $\epsilon>0$.
It is a highly singular function of energy $E$.
This complicates the calculation of the function $\Theta(E)$, and hence,
of the modified Boltzmann entropy $S_\Theta(E)$.
The only obvious solution to this problem is to replace the density
of states by a continuous function. This is of course only meaningful
when the eigenvalues of the Hamiltonian are densely spread over some relevant
interval of energies. A more fundamental solution is lacking and
requires further investigation.


\paragraph*{Acknowledgments}
I thank Prof. D.H.E. Gross for discussions which helped to clarify the text.
Maple 9.5 has been used to produce the drawings.

\end{document}